\documentclass[sigconf]{acmart}

\usepackage{placeins}
\usepackage{booktabs} 

\setcopyright{rightsretained}

\acmDOI{10.475/123_4}

\acmISBN{123-4567-24-567/08/06}

\acmConference[WOODSTOCK'97]{ACM Woodstock conference}{July 1997}{El
  Paso, Texas USA}
\acmYear{1997}
\copyrightyear{2016}

\acmArticle{4}

\begin{document}
\title{Impact of Social Media Posts in Real life Violence: A Case Study in Bangladesh}

\author{Jibon Naher}
\affiliation{%
  \institution{Korea Advanced Institute of Science and Technology}
}
\email{jibon@kaist.ac.kr}

\author{Matiur Rahman Minar}
\affiliation{%
}
\email{minar09.bd@gmail.com}


\begin{abstract}
Social Networking Site (SNS) is a great innovation of modern times. Facebook, Twitter etc. have become an everyday part of peoples' life. Among all SNSs, Facebook is the most popular social network all over the world. Bangladesh is no exception. People of Bangladesh use Facebook for social communication, online shopping, business, knowledge and experience sharing etc. As well as the various uses of SNSs, people sometimes find themselves involved in real life violence, provoked by some social media posts or activities. In this paper, we discussed some case studies in which real life violence is originated based on Facebook activities in Bangladesh. Facebook was used in these incidents intentionally or unintentionally mostly as a tool to trigger hatred and violence. We analyzed and discussed the real-world consequences of these virtual activities in social media. Lastly, we recommended possible future measurements to prevent such violence. 
\end{abstract}

%
%



\maketitle

\section{Introduction}
\label{S:1}
Every single person using internet nowadays uses social networking sites. Social media has become a platform for every kind of communication. Now-a-days one can hardly find anyone who is not a user of any social media, whether an active or silent user. Facebook is the most popular of all the social networking sites. Facebook was originally developed as a communicating site for college students \cite{Facebook:2017wk}. Now, people use Facebook mainly for communicating. Other uses of Facebook include online marketing, knowledge sharing, studying etc. From a report of Statista, in case of monthly active users, Facebook has grown from 100 million to almost 2.2 billion in just one decade. Moreover, approximately, one-third of people use Facebook at least monthly. \cite{Statista2018}. Social media is not only a communicating site now-a-days. People use social media in a variety of ways. There are research works which showed that sometimes usage of social media can be very helpful in personal, educational or professional life\cite{Kim2011TheFP, Gazi2016ProspectiveEL}.       

With the increasing popularity and uses of social media specially Facebook all over the world, people of Bangladesh also find Facebook as a platform for social communication. Currently, there are around 25-30 million Facebook users in Bangladesh \cite{Bdfb:2018bd}.This is the most popular social network site in Bangladesh \cite{Bhabit:2017bd}. Almost all internet users use Facebook regardless of age, gender, locality or ethnic identity \cite{Binternet:2017bd,BinMorshed:2017}. Also, Dhaka is ranked second amontg the top cities for total number of active Facebook users \cite{Bdfb:2018bd}. In Bangladesh, Facebook is now a integral part of daily activities. People use Facebook for social Communication, knowledge and experience sharing, and business purposes and various other reasons \cite{Bhabit:2017bd} .  

Since all sorts of people all over the world are using Facebook for social communications and other purposes, its undeniable that misuses of social networks are bound to happen. Starting from social scams, business frauds, harassment, cheating and many other forms. Almost 57\% women using Facebook have to face online harassment and it is the biggest number of all social networking sites \cite{Statista2017}. Sometimes, women have to face new form of domestic violence, e.g., closing the social media account, because of cyberstalking or harassment in social media \cite{domesticviolence}. Not only women, online harassment is a common phenomena for almost everybody whether he/she is adult or university student or child \cite{harassment:01, harassment:02,harassment:03,harassment:04}. Facebook is used by gangs as a communicating platform \cite{Gangviolence}. Gang violence is also increasing in this era of social networking\cite{King2007SurfAT,Patton2013InternetBN,Patton2016SticksSA}. People are worrying and talking about the increasing youth violence in the social media age \cite{youthviolence}. 

Recently, a new concern is becoming a major issue as well as these negative uses of social media in online. Along with the online harassment, bullying, cheating and gang violence, now-a-days people are severely influenced by social media and take real life actions which is undesirable. Facebook was accused to aid genocide by Myanmar activists. 
\begin{quote}
\textit{clear examples of (Facebook) tools being used to incite real harm.}
\end{quote}
cited by the activits about the incident \cite{Myanmarviolence, Myanmarviolence1}. In India, a Hindu-Muslim violence was triggered by a Facebook post.
\begin{quote}
\textit{A Facebook post was all it took to undo decades of communal harmony in a small east Indian town} \cite{Indiaviolence}.
\end{quote} 
A WhatsApp message in India invoked anger among the people, causing death of seven people from being beaten up \cite{Indiaviolence1}. Mob violence in Sri Lanka was triggered by false stories in Facebook \cite{Sriviolence}. Real life crimes and violences are triggered from social media activities, sharing of images in social media etc. Facebook posts also invoke youths in real life violence and crime, even suiside \cite{Desmond:2014}.   

In Bangladesh, almost all kinds of people using Internet use social media especially Facebook. Thus, similar types of incident are prone to happen. Some terrible incidents occurred in Bangladesh which were triggered by Facebook posts or activities. First major incident such that was in Ramu, Cox's Bazar in 2012. In this study, we are going to briefly discuss about some of the violent incidents triggered by social media in Bangladesh. We will analyze the incidents and lastly recommend some possible preventive measures. We are going to discuss the incidents in section \ref{S:2}. In section \ref{S:3}, we will analyze the cases, discuss the similarities and the links with social media, i.e., Facebook. Section \ref{S:4} is about some possible recommendations about how we can prevent these kind of violence.

\section{Case overview}
\label{S:2}

For the last 5-6 years, quite a number of real life violent incidents occurred in Bangladesh, triggered by some Facebook activities. We will discuss five such incidents where the incidents, were terrible and the loss were tremendous. There are numerous other small incidents which might not get enough attention.

\FloatBarrier    
  \label{tab:incident}
  \begin{tabular}{cccl}
    \toprule
    Number&Year&Location&TriggeredBy\\
    \midrule
    \ 1 & 2012&Ramu&Post image on Facebook\\
    \ 2 & 2013&Pabna&Facebook post\\
    \ 3 & 2014&Comilla&Comment on a Facebook post\\
    \ 4 & 2016 &Brahmanbaria&Facebook post \\
    \ 5 & 2017&Rangpur&Facebook post\\
  \bottomrule
\end{tabular}
\FloatBarrier

In the following subsection, we are going to discuss the cases from one to five. We will analyze the cases from the perspective of how some social media posts and comments can become a terrible tool to trigger real life active violence. 

\subsection{Ramu, 2012}
Ramu is an Upazilla of Cox's Bazar District in Bangladesh. It is located in the  south-east area of the country, under the division of Chittagong. Buddhist temples in Ramu, Cox’s Bazar is of great importance towards Buddhists as well as tourists. \cite{Wikipedia:2017upazilla} \cite{Wikipedia:2017ramu}

In September 2012, Buddhist temples in Ramu, Cox’s Bazar was burnt to the ground. \textbf{This terrible incident was originated from a local Buddhist, Uttam Kumar Barua being tagged in a Facebook image of Quran.} An unknown/fake Facebook user, using a pseudonym, posted burning-Quran image on Uttam Kumar Barua’s Facebook wall. Reacted by the post, a group of arsonists put fire at the Buddhist temple. 
\begin{figure}[hbtp]
\centering
\includegraphics[scale=0.25]{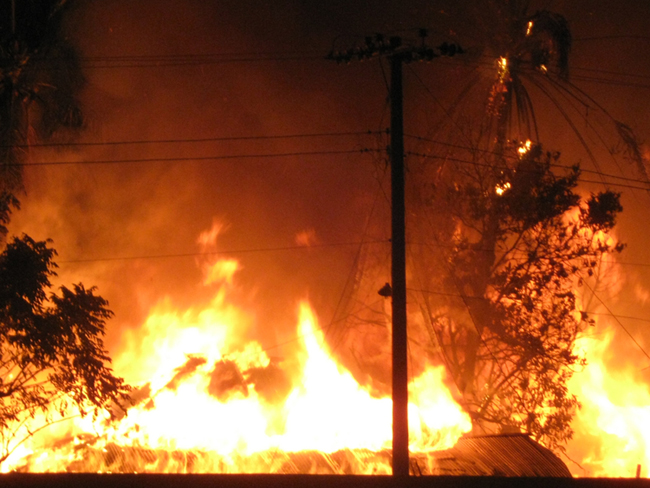} 
\caption{Rampage in Ramu triggered by a Facebook post, 2012}
\label{ramu1}
\end{figure}
Fanatics attacked the Buddhist community in Cox's Bazar's Ramu, claiming that a Buddhist youth “insulted Islam” on social media. \cite{Wikipedia:2012ramu} \cite{DailyStar:2012linked} \cite{BBC:2012facebook} \cite{BdNews:2012ukhiya} \cite{Archive:2012mob} \cite{Archive:2012burned} \cite{DailyStar:2012soul}

\begin{figure}[hbtp]
\centering
\includegraphics[scale=0.26]{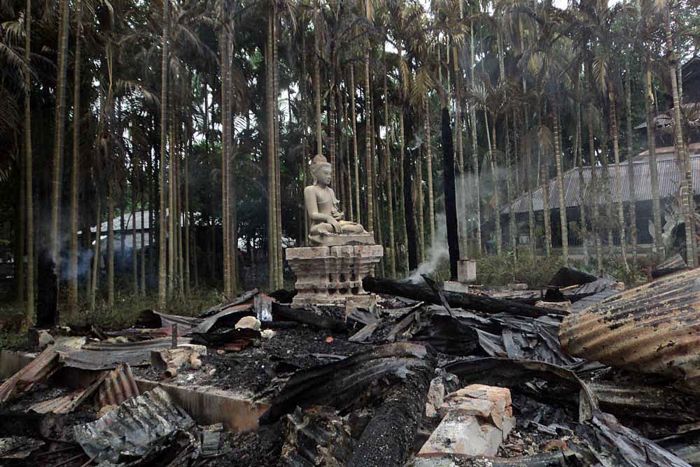}\\
\caption{Destruction of famous Buddhist temples in Ramu incident}
\label{ramu2}
\end{figure}

The mobs destroyed 12 Buddhist temples and monasteries and 50 houses as shown in figure \ref{ramu2}. The violence started in reaction to a tagging of an image depicting the desecration of a Quran on the timeline of a fake Facebook account under a Buddhist male name. The actual posting of the photo was not done by the Buddhist who was falsely slandered.  The Buddhist was innocent of the accusation. The violence later spread to Ukhiya Upazila in Cox's Bazar District and Patiya Upazila in Chittagong District where Buddhist monasteries, Sikh Gurudwaras and Hindu temples were targeted for attacks. \cite{Wikipedia:2012ramu} \cite{DailyStar:2012linked} \cite{BBC:2012facebook} \cite{BdNews:2012ukhiya} \cite{Archive:2012mob} \cite{Archive:2012burned} \cite{HinduExistence:2012ramu}

The man who sparked the riots, who has gone into hiding, told local media he did not post the picture, insisting someone else had \textbf{tagged} his account with the image on the social network \cite{Archive:2012mob}. Later, it was found that, the Facebook page with an anti-Islam picture that provoked the September 29, 2012 rampage against the Buddhist community in Ramu was photoshopped \cite{DailyStar:2012devil}. Somebody or a group had taken a screenshot of Uttam Kumar Barua's facebook profile page, cut out the address of an anti-Islam website and pasted it on the address bar visible in the image. Once the fabrication was done, it looked like the website has shared the anti-Islam image with Uttam and 26 others. \cite{DailyStar:2012devil} \cite{HinduExistence:2012ramu}

\subsection{Pabna, 2013}
\begin{figure}[hbtp]
\centering
\includegraphics[scale=0.28]{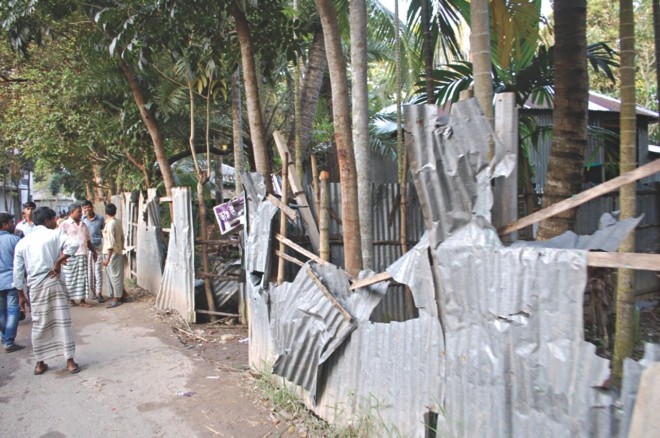} 

\caption{Mob attack in Pabna provoked by a Facebook post in 2013}
\label{pabna}
\end{figure}

This is another occurrence where misuse of Facebook was happened. In November 3, 2013, A mob went on a rampage in a Hindu-dominated neighbourhood in Bonogram of Pabna, Bangladesh following reports that a boy from the minority community had committed blasphemy. As a result, more than 25 houses belonging to Hindus vandalised as shown in figure \ref{pabna}, several idols in temples damaged and about 150 families forced to flee the area  \cite{Persecution:2013mob} \cite{HinduExistence:2013pabna} \cite{DailyStar:2013attacked}.

A group of people were distributing photocopies of what they said was a “Facebook page”. They claimed one Rajib Saha had maligned Prophet Mohammad (pbuh) in the page \cite{DailyStar:2013attacked} \cite{HinduHumanRights:2013false}. An eyewitness said,
\begin{quote}
\textit{None was given the chance to ask whether or not it was a faked Facebook posting.}\cite{DailyStar:2013attacked}
\end{quote}
Rajib, son of Babul Saha, a shop owner in the bazaar, is a class-X student of Bonogram Miapur High School \cite{DailyStar:2013attacked}. Babul Saha said,
\begin{quote}
\textit{He was preparing for SSC examination. He can't do anything like what the people here are alleging.}\cite{DailyStar:2013attacked}
\end{quote}

The name of the Facebook page was written in Bangla and denigrates the prophet. It was opened on September 14, 2013. It did contain hateful posts, majority of which was issued by the administrator. Some people gave negative reactions to these posts, asking others to refrain from liking this page \cite{DailyStar:2013attacked} \cite{HinduHumanRights:2013false}. It was later found that the Facebook page whose photocopies were used to incite the attacks on the Bonogram Hindu community has no links with Rajib who was being accussed \cite{DailyStar:2013attacked} \cite{HinduExistence:2013pabna}.

\subsection{Comilla, 2014}
\begin{figure}[hbtp]
\centering
\includegraphics[scale=0.3]{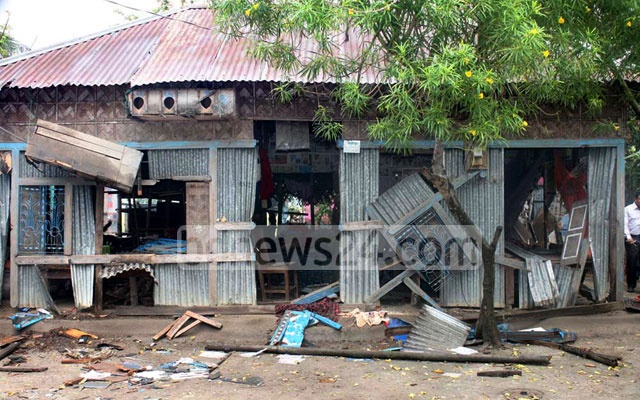}
\caption{Attack triggered by Facebook comment in Comilla, 2014}
\label{comilla}
\end{figure}

\textbf{Fake news of defaming religion on Facebook triggered violent attacks in Comilla, 2014.} It was in Homna Upazilla of Comilla, Bangladesh over Hindu minority \cite{HinduExistence:2014comilla} as shown in figure \ref{comilla}.

At least 28 houses of Hindus were ransacked at Bakhsitarampur village in Homna upazila of Comilla in an attack of a mob of nearly 3,000 prompted by rumours that Prophet Muhammad (pbuh) had been defamed in Facebook posts by some Hindus \cite{DailyStar:2014rumour} \cite{BdChronicle:2014attack} \cite{HinduSamhati:2014smashed} \cite{HinduExistence:2014comilla}. Homna police station OC Aslam Shikder said that hundreds of people from the Panchkipta village attacked temples and homes of at least 28 Hindu families alleging that two Hindu youths posted defamatory comments on Facebook about the Prophet \cite{BdChronicle:2014attack}. Police detained two persons Utshab Das and Srinibas Das as well as nine others over the rumours of Facebook post  \cite{BdChronicle:2014attack}. Homna police station OC Aslam Shikder said,
\begin{quote}
\textit{During the interrogation they have denied posting any such remark.} \cite{BdChronicle:2014attack}
\end{quote}

Villagers said a call was made from the loudspeakers of a Madrasa at Rampur village, near Baghsitarampur, to launch the assault on the Hindus. Before the attacks, leaflets were distributed for last several days in the madrasas claiming that two Hindu youths had slandered the prophet in a Facebook post on April 27, 2014. \cite{BdNews:2014madrasa}

\subsection{Brahmanbaria, 2016}
\begin{figure}[hbtp]
\includegraphics[scale=0.35]{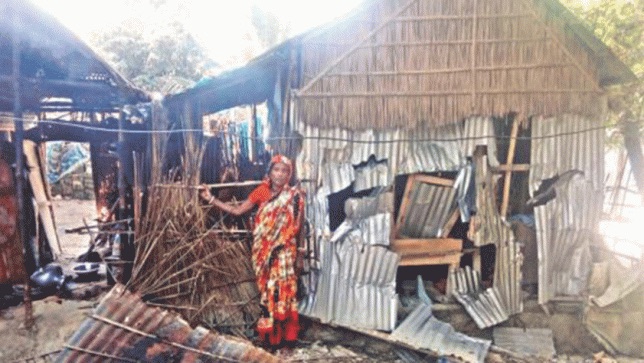}
\caption{Brahmanbaria attack provoked by a Facebook post, 2016}
\label{Bbaria}
\end{figure}

In October, 2016, violent mob carried out a synchronized attack on the Hindus in Brahmanbaria's Nasirnagar upazil over an alleged Facebook post. Destroying and setting fire on more than 150 homes and at least 15 temples and looting valuables \cite{HinduExistence:2016bbaria} provoked by the Facebook post insulting Islam \cite{BdNews:2016withdrawn} \cite{DailyStar:2016police}, as shown in figure \ref{Bbaria}. At least 20 people including several temple devotees were left wounded in the attack \cite{BdNews:2016withdrawn}. Later, twice more, attacks were carried out on Hindus – setting their houses on fire \cite{DailyStar:2016police} \cite{HinduExistence:2016bbaria} \cite{DailyStar:2016planned}.

The violence was triggered by a Facebook post purportedly from the account named 'Rasraj Das', son of Jagannath Das at Haripur Union’s Harinberh village for "hurting religious sentiments of Muslims", as the locals said \cite{BdNews:2016withdrawn}. Police later arrested Rasraj for ‘denigrating Islam’ through his post on the social media. A court then ordered him into prison \cite{BdNews:2016withdrawn} \cite{DailyStar:2016suspect} \cite{Breakingnews:2017two} \cite{DailyStar:2016planned}.

It was later exposed that Awami League leader Faruk Mia, the District Union President of Nasirnagar had some problem with the local Fishermen Union leader Rasaraj Das \cite{DailyStar:2016planned}. Hence, Faruk opened an Facebook account in the name of Rasaraj Das \cite{HinduExistence:2016bbaria}.Then, Faruk posted a picture of Kaba juxtaposed with Hindu deity Lord Shiva with the help of his brother Kaptan Mia in the timeline of Rasaraj \cite{HinduExistence:2016bbaria}.

\subsection{Rangpur, 2017} 
\begin{figure}[hbtp]
\includegraphics[scale=0.2]{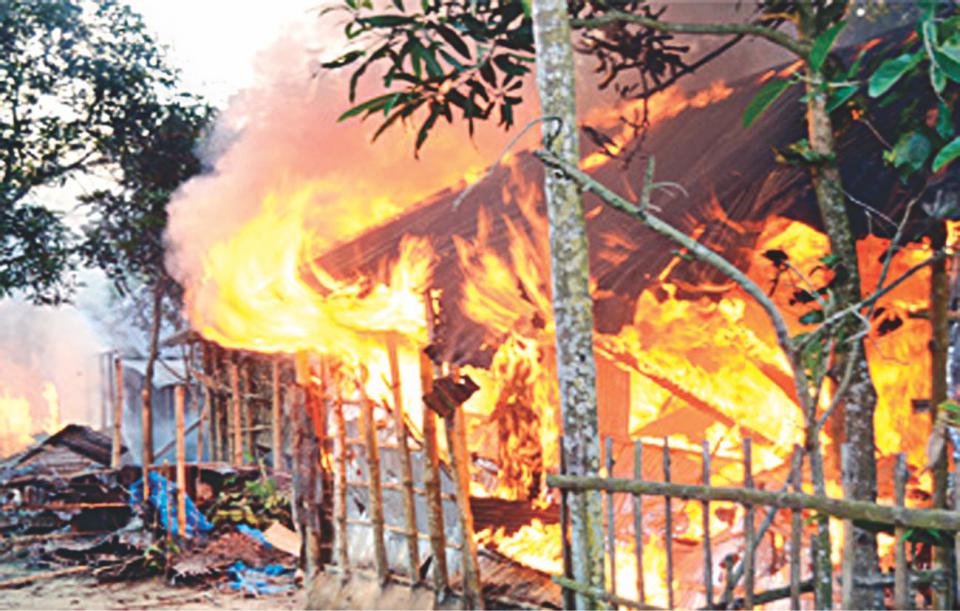}\\
\caption{Rangpur incident triggered by Facebook post, 2017}
\label{Rangpur1}
\end{figure}

In November 10, 2017, a clash broke out in Thakurpara, Rangpur \cite{DailyStar:2017clash} which was triggered by a Facebook post \cite{DailyStar:2017mayhem}. This resulted 1 killed, 20 hurt including 7 policemen \cite{DailyStar:2017clash} \cite{DailyStar:2017mayhem} \cite{BdNews:2017saved}. At least 30 Hindu houses were burned and vandalized in Horkoli Thakurpara village of Rangpur \cite{DailyStar:2017mayhem} \cite{BdNews:2017saved} as shown in figure \ref{Rangpur1}.

This incident was triggered by a Facebook post. In November 5, 2017, Titu Chandra Roy from Rangpur, Bangladesh who was recently living in Narayanganj, Bangladesh shared a Facebook status said to be “Defaming religion” \cite{DailyStar:2017mayhem}. Similar to Ramu violence in 2012 \cite{Wikipedia:2012ramu}, a group of arsonists put fire to the property of Hindu minority at Horkoli Thakurpara village of Rangpur \cite{DailyStar:2017mayhem}. The controversial Facebook post being ‘defamation of Prophet Muhammad’, triggered the communal attack on the Hindu houses by irrate mob in Thakurpara village of Rangpur \cite{HinduExistence:2017rangpur}. The Facebook account resembling Titu Roy is named as MD Titu. The account was opened just two months before of the incident, in September 2017, although it had managed to make 288 Facebook friends by this time \cite{HinduExistence:2017rangpur}.
Ziton Bala, mother of Tito Chandra Roy, said \cite{DailyStar:2017framed} \cite{DailyStar:2017accused}, 
\begin{quote}
\textit{My son's name is Tito; people are telling me his name is Titu and it's not true. }
\end{quote}
The home minister said \cite{BdNews:2017arrested}, 
\begin{quote}
\textit{We have heard that Titu is illiterate }\cite{HinduExistence:2017rangpur}.
\end{quote}

\begin{figure}[hbtp]
\includegraphics[scale=0.17]{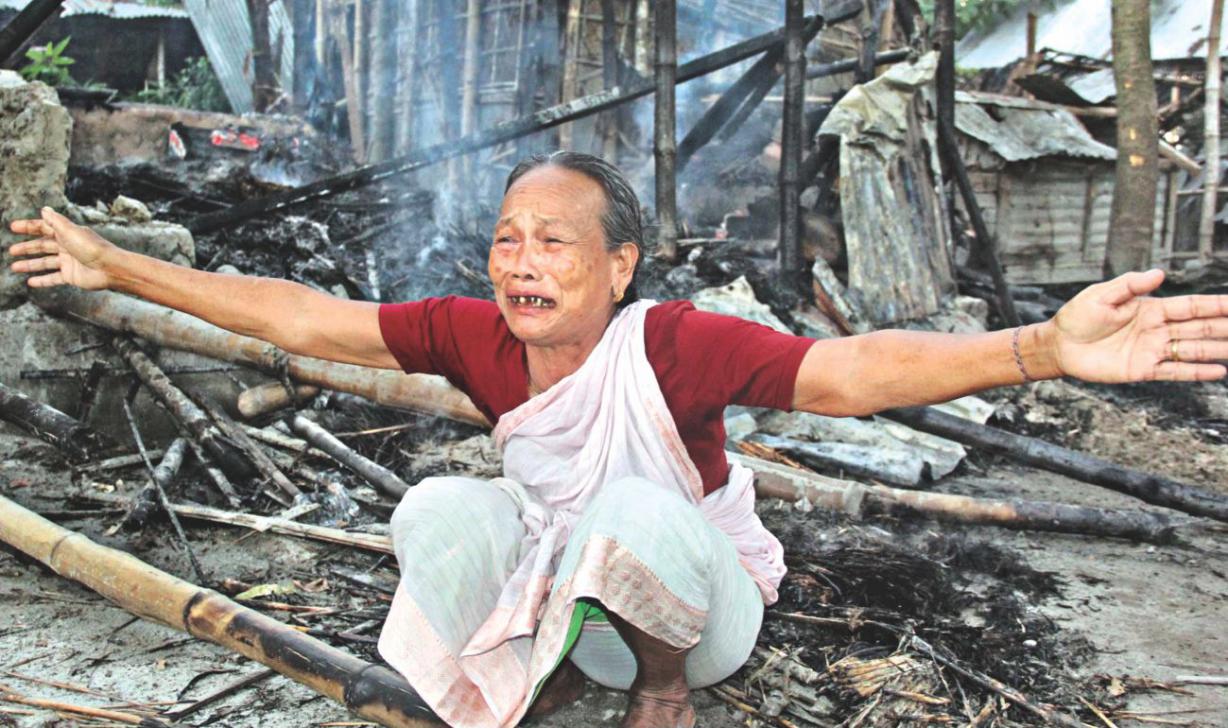}
\caption{Mayham in Rangpur, 2017}
\label{Rangpur2}
\end{figure}

Sirajul Islam, an Imam of a mosque in a nearby village, and Alamgir Hossain, a trader, filed a case against one Titu Chandra Roy with Gangachara Police Station, accusing him of making a post on Facebook that hurt religious sentiments of Muslims \cite{DailyStar:2017framed} \cite{DailyStar:2017accused}. Then the influential locals spread hatred which resulted the violence and clash with police \cite{DailyStar:2017framed}. A mobile phone number is found which was used for opening the Facebook ID around two months ago. The phone number was found unused and police could not verify who the number was registered to. The owner of the Facebook account generally shared posts or uploaded screenshots of others' posts \cite{DailyStar:2017framed} \cite{DailyStar:2017accused}. Tito Chandra Roy, an accused in a case filed for allegedly making a post on Facebook that sparked mayhem in Thakurpara village of Rangpur, was arrested in Jaldhaka upazila of Nilphamari district in November 14, 2017 \cite{DailyStar:2017accused} \cite{BdNews:2017arrested}. 
Rangpur Superintendent of Police Mizanur Rahman said  \cite{BdNews:2017arrested},
\begin{quote}
\textit{We did not find any status insulting religion on his Facebook. False information was spread to stir attacks, arson and looting against the Hindu community.}
\end{quote}

\section{Analysis and Discussions}
\label{S:3}
We are now going to analyze deeply about how the incidents triggered at the first place from social media with respect to our discussion in \ref{S:2}. We will try to find the link, relationship and dependencies of social media in all the cases.
\subsection{Ramu}
In the Ramu incident, from the reports and articles it is clear that the incident was triggered intentionally. The accused was framed by someone by being tagged in a fake Facebook image. Facebook was used there as a medium of reaching a large number of people. Nobody tried to question the authenticity of the Facebook post. Someone took the advantage of social media and a violence occurred.  
\subsection{Pabna}
A boy was accused in this case to commit blasphemy in a Facebook page against a religion. However, it was later found that the boy had no connection with the page. He was framed by someone using the popularity of Facebook. 
\subsection{Comilla} 
This incident was triggered by defaming the accused about commenting something objectionable on a Facebook post. Which was said to be fake later. There was not any clear evidence that the accused actually did some comments which were defamatory. 
\subsection{Brahmanbaria}  
In this incident, someone used Facebook to get revenge regarding personal issue. He opened a Facebook account in the name of another person and posted something questionable, which ignited the people. That caused a mob attack. 
\subsection{Rangpur}
Almost same type of misuse of Facebook was happened in Rangpur too. The accused was said to be allegedly making a post on Facebook that sparked mayhem in Thakurpara village of Rangpur. However, none found any proof that he actually did something like that.          

In this paper, we discussed five of the caeses where Facebook post or comments or images were used to provoke general people into violence and mob attack. Bangladesh saw dozen of virulent attacks on minorities with spreading rumors or posting objectionable elements on social media similar to this. In all these cases, perpetrators took the advantage of popular social media platform Facebook to generate fundamental sentiments to attack minorities in Bangladesh \cite{HinduExistence:2016bbaria} \cite{HinduExistence:2017rangpur}.

From these cases, one thing is common. All started from Facebook posts or comments or sharing images in Facebook, which triggered arson and violence. Also, the arsonists and attackers seem to be greatly affected by social network i.e. Facebook posts or activities. This indicates the availability and expansion of social networks in a third-world and developing country like Bangladesh. This also indicates the ignorance and unwillingness of local or related people towards verification of authenticity of such defaming posts or activities. 

Facebook activities are seem to be taken into granted here without any context or authenticity. Also it seems some are taking advantages of this ignorance, trying deliberately to drive people towards hatred and extremism, while gaining different purposes.

\section{Recommendation for future prevention}
\label{S:4}
Many research and development works are being done using technology oriented solutions for Bangladeshi social problems \cite{Ahmed:2015:RMI:2702123.2702573}. For example, Protibadi \cite{Ahmed:2014:PPF:2611205.2557376} is a notable one. Also there are works like remote health monitoring \cite{Minar:2017vital} in the context of Bangladesh. \cite{MBGDMQN2013} studied some trust issues of social networks and proposed ways to solve them.

After discussing and analyzing the case studies, we think that there are some possible measurements that can be taken to prevent this kind of active violence based on social media, specially in Bangladesh.
\subsection{Technological perspective}
The violence were provoked from social media, specially Facebook. From technological perspective, there may be some steps to be taken so that prevention is possible for such mob violence. The following are possible recommendations that can be taken, in the context of Bangladesh-  
\begin{itemize}
  \item Considering this study, social networks need to be stricter on moderating users’ identity and activities. It seems that some people are deliberately using Facebook as a medium of triggering violence. Social media can use multiple levels of identity verification of the users and their purposes in social networks and take suitable steps to identify fake/inactive users and close their accounts as well.
  \item Detection of harassment, bullying in social media can be useful in preventing violence. There are already existing works for detecting online harassment or cyberbulling \cite{W18-4418,W18-4417}. In some of the cases we discussed, Facebook was used for religious or personal revenge. Thus, detecting posts or comments in social media related to religion can be useful in preventing such violence in a country like Bangladesh.      
  \item To prevent these kind of violence beforehand, social networks and defensive forces can use detection system over Facebook and other social networks. Detecting suspicious users and activities from social networks could be useful in this regard. For example, Facebook recently launched Artificial Intelligence based program to predict and prevent suicides \cite{Wired:2017suicide}. Similar programs can be implemented to predict and prevent violence and extremism. 
\end{itemize}
\subsection{Psychological perspective} 
In all of the cases, people were greatly influenced by social media. It is very interesting and a matter of study how the virtual world provoked someone in the extent to take physical action, e.g., mob violence. So, one can not but to agree to take some steps about the use of social media from psychological perspective.  
\begin{itemize}
  \item Now-a-days everybody is dependent on online social media. Social network influence peoples' life in almost every way. Every area of life whether it is lifestyle, academic or business, virtual media plays an important role \cite{Alwagait:2015:ISM:2828640.2828765, Paul:2012:EOS:2365365.2365675}. It is like people cannot no more differentiate properly between virtual and physical life. People are being more and more involved in virtual world leaving the physical world. We think people should increase real life activities with friends and family. It will help them to not engaged so much in virtual world and influence negatively. 
  \item In all the cases we discussed in this paper, people believed social media blindly. Nobody questioned about the authenticity of the Facebook posts or tried to prove whether the news is fake or not. People should not trust virtual world at this level because it is very easy to frame someone in this social media age. Which happened in some of the cases actually.    
\end{itemize}

\section{Conclusion}
The case studies discussed in this paper give an overview about the alarming phenomenon of real life violence provoked by Facebook activites in Bangladesh. However, in most of the cases, the post was fake or someone intentionally misuse the popularity of social media for personal reason. It is a matter of great anxiety that such violence occurred in the first place. Social media is a very useful place for people to reach, communicate and contact lots of people at the same time. There should be some policies that include guidelines for online communications established by some respective authority. Well defined infrastructure may be developed that promotes safety, and provide intervention in using social media when needed. Comprehensive research efforts to define, detect and examine such active violence provoked by social media are needed.

\bibliographystyle{ACM-Reference-Format}
\bibliography{Arsonisms-sample-bibliography}


\begin{thebibliography}{67}


\ifx \showCODEN    \undefined \def \showCODEN     #1{\unskip}     \fi
\ifx \showDOI      \undefined \def \showDOI       #1{#1}\fi
\ifx \showISBNx    \undefined \def \showISBNx     #1{\unskip}     \fi
\ifx \showISBNxiii \undefined \def \showISBNxiii  #1{\unskip}     \fi
\ifx \showISSN     \undefined \def \showISSN      #1{\unskip}     \fi
\ifx \showLCCN     \undefined \def \showLCCN      #1{\unskip}     \fi
\ifx \shownote     \undefined \def \shownote      #1{#1}          \fi
\ifx \showarticletitle \undefined \def \showarticletitle #1{#1}   \fi
\ifx \showURL      \undefined \def \showURL       {\relax}        \fi
\providecommand\bibfield[2]{#2}
\providecommand\bibinfo[2]{#2}
\providecommand\natexlab[1]{#1}
\providecommand\showeprint[2][]{arXiv:#2}

\bibitem[\protect\citeauthoryear{QUARTZINDIA}{Ind}{2017b}]%
        {Indiaviolence}
 \bibinfo{year}{2017}\natexlab{b}.
\newblock \bibinfo{title}{A Facebook post was all it took to undo decades of
  communal harmony in a small east Indian town}.
\newblock
\newblock
\urldef\tempurl%
\url{https://bit.ly/2C0qJsx}
\showURL{%
Retrieved August 28, 2018 from \tempurl}


\bibitem[\protect\citeauthoryear{hindustantimes}{Ind}{2017a}]%
        {Indiaviolence1}
 \bibinfo{year}{2017}\natexlab{a}.
\newblock \bibinfo{title}{Jharkhand lynching: When a WhatsApp message turned
  tribals into killer mobs}.
\newblock
\newblock
\urldef\tempurl%
\url{https://bit.ly/2LEUX3H}
\showURL{%
Retrieved August 28, 2018 from \tempurl}


\bibitem[\protect\citeauthoryear{Fortune}{Sri}{2018}]%
        {Sriviolence}
 \bibinfo{year}{2018}\natexlab{}.
\newblock \bibinfo{title}{Facebook Accused of Ignoring Government Warnings
  Before Mob Violence in Sri Lanka}.
\newblock
\newblock
\urldef\tempurl%
\url{http://fortune.com/2018/04/22/facebook-ignored-sri-lanka-hate-speech/}
\showURL{%
Retrieved August 28, 2018 from \tempurl}


\bibitem[\protect\citeauthoryear{aljazeera}{Mya}{2018a}]%
        {Myanmarviolence1}
 \bibinfo{year}{2018}\natexlab{a}.
\newblock \bibinfo{title}{UN: Facebook had a 'role' in Rohingya genocide}.
\newblock
\newblock
\urldef\tempurl%
\url{https://bit.ly/2BWzX92}
\showURL{%
Retrieved August 28, 2018 from \tempurl}


\bibitem[\protect\citeauthoryear{CNN}{Mya}{2018b}]%
        {Myanmarviolence}
 \bibinfo{year}{2018}\natexlab{b}.
\newblock \bibinfo{title}{When Facebook becomes 'the beast': Myanmar activists
  say social media aids genocide}.
\newblock
\newblock
\urldef\tempurl%
\url{https://cnn.it/2Nu1pfW}
\showURL{%
Retrieved August 28, 2018 from \tempurl}


\bibitem[\protect\citeauthoryear{Ahmed, Jackson, Ahmed, Ferdous, Rifat, Rizvi,
  Ahmed, and Mansur}{Ahmed et~al\mbox{.}}{2014}]%
        {Ahmed:2014:PPF:2611205.2557376}
\bibfield{author}{\bibinfo{person}{Syed~Ishtiaque Ahmed},
  \bibinfo{person}{Steven~J. Jackson}, \bibinfo{person}{Nova Ahmed},
  \bibinfo{person}{Hasan~Shahid Ferdous}, \bibinfo{person}{Md.~Rashidujjaman
  Rifat}, \bibinfo{person}{A.S.M Rizvi}, \bibinfo{person}{Shamir Ahmed}, {and}
  \bibinfo{person}{Rifat~Sabbir Mansur}.} \bibinfo{year}{2014}\natexlab{}.
\newblock \showarticletitle{Protibadi: A Platform for Fighting Sexual
  Harassment in Urban Bangladesh}. In \bibinfo{booktitle}{\emph{Proceedings of
  the 32Nd Annual ACM Conference on Human Factors in Computing Systems}}
  \emph{(\bibinfo{series}{CHI '14})}. \bibinfo{publisher}{ACM},
  \bibinfo{address}{New York, NY, USA}, \bibinfo{pages}{2695--2704}.
\newblock
\showISBNx{978-1-4503-2473-1}
\urldef\tempurl%
\url{https://doi.org/10.1145/2556288.2557376}
\showDOI{\tempurl}


\bibitem[\protect\citeauthoryear{Ahmed, Mim, and Jackson}{Ahmed
  et~al\mbox{.}}{2015}]%
        {Ahmed:2015:RMI:2702123.2702573}
\bibfield{author}{\bibinfo{person}{Syed~Ishtiaque Ahmed},
  \bibinfo{person}{Nusrat~Jahan Mim}, {and} \bibinfo{person}{Steven~J.
  Jackson}.} \bibinfo{year}{2015}\natexlab{}.
\newblock \showarticletitle{Residual Mobilities: Infrastructural Displacement
  and Post-Colonial Computing in Bangladesh}. In
  \bibinfo{booktitle}{\emph{Proceedings of the 33rd Annual ACM Conference on
  Human Factors in Computing Systems}} \emph{(\bibinfo{series}{CHI '15})}.
  \bibinfo{publisher}{ACM}, \bibinfo{address}{New York, NY, USA},
  \bibinfo{pages}{437--446}.
\newblock
\showISBNx{978-1-4503-3145-6}
\urldef\tempurl%
\url{https://doi.org/10.1145/2702123.2702573}
\showDOI{\tempurl}


\bibitem[\protect\citeauthoryear{Alwagait, Shahzad, and Alim}{Alwagait
  et~al\mbox{.}}{2015}]%
        {Alwagait:2015:ISM:2828640.2828765}
\bibfield{author}{\bibinfo{person}{Esam Alwagait}, \bibinfo{person}{Basit
  Shahzad}, {and} \bibinfo{person}{Sophia Alim}.}
  \bibinfo{year}{2015}\natexlab{}.
\newblock \showarticletitle{Impact of Social Media Usage on Students Academic
  Performance in Saudi Arabia}.
\newblock \bibinfo{journal}{\emph{Comput. Hum. Behav.}} \bibinfo{volume}{51},
  \bibinfo{number}{PB} (\bibinfo{date}{Oct.} \bibinfo{year}{2015}),
  \bibinfo{pages}{1092--1097}.
\newblock
\showISSN{0747-5632}
\urldef\tempurl%
\url{https://doi.org/10.1016/j.chb.2014.09.028}
\showDOI{\tempurl}


\bibitem[\protect\citeauthoryear{Archive}{Archive}{2012}]%
        {Archive:2012mob}
\bibfield{author}{\bibinfo{person}{Archive}.} \bibinfo{year}{2012}\natexlab{}.
\newblock \bibinfo{title}{Rioting mob torches temples in Bangladesh}.
\newblock
\newblock
\urldef\tempurl%
\url{https://bit.ly/2LDHz00}
\showURL{%
\tempurl}


\bibitem[\protect\citeauthoryear{Arroyo-Fern{\'a}ndez, Forest, Torres-Moreno,
  Carrasco-Ruiz, Legeleux, and Joannette}{Arroyo-Fern{\'a}ndez
  et~al\mbox{.}}{2018}]%
        {W18-4417}
\bibfield{author}{\bibinfo{person}{Ignacio Arroyo-Fern{\'a}ndez},
  \bibinfo{person}{Dominic Forest}, \bibinfo{person}{Juan-Manuel
  Torres-Moreno}, \bibinfo{person}{Mauricio Carrasco-Ruiz},
  \bibinfo{person}{Thomas Legeleux}, {and} \bibinfo{person}{Karen Joannette}.}
  \bibinfo{year}{2018}\natexlab{}.
\newblock \showarticletitle{Cyberbullying Detection Task: the EBSI-LIA-UNAM
  System (ELU) at COLING'18 TRAC-1}. In \bibinfo{booktitle}{\emph{Proceedings
  of the First Workshop on Trolling, Aggression and Cyberbullying
  (TRAC-2018)}}. \bibinfo{publisher}{Association for Computational
  Linguistics}, \bibinfo{pages}{140--149}.
\newblock
\urldef\tempurl%
\url{http://aclweb.org/anthology/W18-4417}
\showURL{%
\tempurl}


\bibitem[\protect\citeauthoryear{bangladeshchronicle.net}{bangladeshchronicle.net}{2014}]%
        {BdChronicle:2014attack}
\bibfield{author}{\bibinfo{person}{bangladeshchronicle.net}.}
  \bibinfo{year}{2014}\natexlab{}.
\newblock \bibinfo{title}{Rumour triggers attack on Hindus}.
\newblock
\newblock
\urldef\tempurl%
\url{http://bangladeshchronicle.net/2014/04/rumour-triggers-attack-on-hindus/}
\showURL{%
\tempurl}


\bibitem[\protect\citeauthoryear{bbc.com}{bbc.com}{2012}]%
        {BBC:2012facebook}
\bibfield{author}{\bibinfo{person}{bbc.com}.} \bibinfo{year}{2012}\natexlab{}.
\newblock \bibinfo{title}{Bangladesh rampage over Facebook Koran image}.
\newblock
\newblock
\urldef\tempurl%
\url{http://www.bbc.com/news/world-asia-19780692}
\showURL{%
\tempurl}


\bibitem[\protect\citeauthoryear{bdnews24.com}{bdnews24.com}{2012a}]%
        {BdNews:2012ukhiya}
\bibfield{author}{\bibinfo{person}{bdnews24.com}.}
  \bibinfo{year}{2012}\natexlab{a}.
\newblock \bibinfo{title}{5 Buddhist temples attacked in Ukhia}.
\newblock
\newblock
\urldef\tempurl%
\url{https://bdnews24.com/bangladesh/2012/09/30/5-buddhist-temples-attacked-in-ukhia}
\showURL{%
\tempurl}


\bibitem[\protect\citeauthoryear{bdnews24.com}{bdnews24.com}{2012b}]%
        {Archive:2012burned}
\bibfield{author}{\bibinfo{person}{bdnews24.com}.}
  \bibinfo{year}{2012}\natexlab{b}.
\newblock \bibinfo{title}{Buddhist temples, homes burned, looted in Ramu}.
\newblock
\newblock
\urldef\tempurl%
\url{https://bit.ly/2C05aIr}
\showURL{%
\tempurl}


\bibitem[\protect\citeauthoryear{bdnews24.com}{bdnews24.com}{2014}]%
        {BdNews:2014madrasa}
\bibfield{author}{\bibinfo{person}{bdnews24.com}.}
  \bibinfo{year}{2014}\natexlab{}.
\newblock \bibinfo{title}{Madrasa people led attacks on Hindus}.
\newblock
\newblock
\urldef\tempurl%
\url{https://bdnews24.com/bangladesh/2014/05/05/madrasa-people-led-attacks-on-hindus}
\showURL{%
\tempurl}


\bibitem[\protect\citeauthoryear{bdnews24.com}{bdnews24.com}{2016}]%
        {BdNews:2016withdrawn}
\bibfield{author}{\bibinfo{person}{bdnews24.com}.}
  \bibinfo{year}{2016}\natexlab{}.
\newblock \bibinfo{title}{Nasirnagar police OC withdrawn for negligence during
  Brahmanbarhia temple attack}.
\newblock
\newblock
\urldef\tempurl%
\url{https://bit.ly/2wxeEFk}
\showURL{%
\tempurl}


\bibitem[\protect\citeauthoryear{bdnews24.com}{bdnews24.com}{2017a}]%
        {BdNews:2017saved}
\bibfield{author}{\bibinfo{person}{bdnews24.com}.}
  \bibinfo{year}{2017}\natexlab{a}.
\newblock \bibinfo{title}{Hindus' homes in Rangpur could not be saved as mob
  outgrew police, home minister says}.
\newblock
\newblock
\urldef\tempurl%
\url{https://bit.ly/2MGPIGt}
\showURL{%
\tempurl}


\bibitem[\protect\citeauthoryear{bdnews24.com}{bdnews24.com}{2017b}]%
        {BdNews:2017arrested}
\bibfield{author}{\bibinfo{person}{bdnews24.com}.}
  \bibinfo{year}{2017}\natexlab{b}.
\newblock \bibinfo{title}{Man arrested for ‘blasphemous’ Facebook post that
  sparked Rangpur arson}.
\newblock
\newblock
\urldef\tempurl%
\url{https://bit.ly/2N36fnh}
\showURL{%
\tempurl}


\bibitem[\protect\citeauthoryear{Bin~Morshed, Dye, Ahmed, and
  Kumar}{Bin~Morshed et~al\mbox{.}}{2017}]%
        {BinMorshed:2017}
\bibfield{author}{\bibinfo{person}{Mehrab Bin~Morshed},
  \bibinfo{person}{Michaelanne Dye}, \bibinfo{person}{Syed~Ishtiaque Ahmed},
  {and} \bibinfo{person}{Neha Kumar}.} \bibinfo{year}{2017}\natexlab{}.
\newblock \showarticletitle{When the Internet Goes Down in Bangladesh}. In
  \bibinfo{booktitle}{\emph{Proceedings of the 2017 ACM Conference on Computer
  Supported Cooperative Work and Social Computing}}
  \emph{(\bibinfo{series}{CSCW '17})}. \bibinfo{publisher}{ACM},
  \bibinfo{address}{New York, NY, USA}, \bibinfo{pages}{1591--1604}.
\newblock
\urldef\tempurl%
\url{https://doi.org/10.1145/2998181.2998237}
\showDOI{\tempurl}


\bibitem[\protect\citeauthoryear{breakingnews.com.bd}{breakingnews.com.bd}{2017}]%
        {Breakingnews:2017two}
\bibfield{author}{\bibinfo{person}{breakingnews.com.bd}.}
  \bibinfo{year}{2017}\natexlab{}.
\newblock \bibinfo{title}{Two detained in Nasirnagar, Brahmanbaraia}.
\newblock
\newblock
\urldef\tempurl%
\url{http://www.breakingnews.com.bd/bn/english/news/7859}
\showURL{%
\tempurl}


\bibitem[\protect\citeauthoryear{businesshabit.com}{businesshabit.com}{2015}]%
        {Bhabit:2017bd}
\bibfield{author}{\bibinfo{person}{businesshabit.com}.}
  \bibinfo{year}{2015}\natexlab{}.
\newblock \bibinfo{title}{The Number of Facebook Users in Bangladesh}.
\newblock
\newblock
\urldef\tempurl%
\url{https://www.businesshabit.com/2015/10/the-number-of-facebook-users-in.html}
\showURL{%
Retrieved August 27, 2018 from \tempurl}


\bibitem[\protect\citeauthoryear{Desmond Upton~Patton}{Desmond
  Upton~Patton}{2014}]%
        {Desmond:2014}
\bibfield{author}{\bibinfo{person}{Megan Ranney Sadiq Patel Caitlin Kelley Rob
  Eschmann Tyreasa~Washington Desmond Upton~Patton, Jun Sung~Hong}.}
  \bibinfo{year}{2014}\natexlab{}.
\newblock \showarticletitle{Social media as a vector for youth violence: A
  review of the literature}.
\newblock \bibinfo{journal}{\emph{Elsevier}}  \bibinfo{volume}{35}
  (\bibinfo{date}{June} \bibinfo{year}{2014}), \bibinfo{pages}{548--553}.
\newblock
\urldef\tempurl%
\url{https://doi.org/10.1016/j.chb.2014.02.043}
\showDOI{\tempurl}


\bibitem[\protect\citeauthoryear{Digiology}{Digiology}{2018}]%
        {Bdfb:2018bd}
\bibfield{author}{\bibinfo{person}{Digiology}.}
  \bibinfo{year}{2018}\natexlab{}.
\newblock \bibinfo{title}{Demographics of Facebook Population in Bangladesh,
  April 2018}.
\newblock
\newblock
\urldef\tempurl%
\url{http://digiology.xyz/demographics-facebook-population-bangladesh-april-2018/}
\showURL{%
Retrieved August 27, 2018 from \tempurl}


\bibitem[\protect\citeauthoryear{Faye~Mishna}{Faye~Mishna}{2009}]%
        {harassment:03}
\bibfield{author}{\bibinfo{person}{Steven~Solomon Faye~Mishna, Michael~Saini}.}
  \bibinfo{year}{2009}\natexlab{}.
\newblock \showarticletitle{Ongoing and online: Children and youth's
  perceptions of cyber bullying}.
\newblock \bibinfo{journal}{\emph{Elsevier}}  \bibinfo{volume}{31}
  (\bibinfo{date}{December} \bibinfo{year}{2009}), \bibinfo{pages}{1222--1228}.
\newblock
Issue 12.
\urldef\tempurl%
\url{https://doi.org/10.1016/j.childyouth.2009.05.004}
\showDOI{\tempurl}


\bibitem[\protect\citeauthoryear{Finn}{Finn}{2004}]%
        {harassment:02}
\bibfield{author}{\bibinfo{person}{Jerry Finn}.}
  \bibinfo{year}{2004}\natexlab{}.
\newblock \showarticletitle{A Survey of Online Harassment at a University
  Campus}.
\newblock \bibinfo{journal}{\emph{Interpersonal Violence}}
  \bibinfo{volume}{19} (\bibinfo{year}{2004}).
\newblock
Issue 4.
\urldef\tempurl%
\url{https://doi.org/10.1177/0886260503262083}
\showDOI{\tempurl}


\bibitem[\protect\citeauthoryear{futurestartup.com}{futurestartup.com}{2017}]%
        {Binternet:2017bd}
\bibfield{author}{\bibinfo{person}{futurestartup.com}.}
  \bibinfo{year}{2017}\natexlab{}.
\newblock \bibinfo{title}{Facebook Rules The Internet In Bangladesh and Many
  Users Don’t Know That They Are Using The Internet}.
\newblock
\newblock
\urldef\tempurl%
\url{tiny.cc/t53nxy}
\showURL{%
Retrieved August 28, 2018 from \tempurl}


\bibitem[\protect\citeauthoryear{Gazi}{Gazi}{2016}]%
        {Gazi2016ProspectiveEL}
\bibfield{author}{\bibinfo{person}{Cem~Balcikanli Gazi}.}
  \bibinfo{year}{2016}\natexlab{}.
\newblock \showarticletitle{Prospective English language teachers ’
  experiences in Facebook : Adoption , use and educational use in Turkish
  context}.
\newblock


\bibitem[\protect\citeauthoryear{Guardian}{Guardian}{2018}]%
        {youthviolence}
\bibfield{author}{\bibinfo{person}{The Guardian}.}
  \bibinfo{year}{2018}\natexlab{}.
\newblock \bibinfo{title}{Social media related to violence by young people, say
  experts}.
\newblock
\newblock
\urldef\tempurl%
\url{https://bit.ly/2GzSlCw}
\showURL{%
Retrieved August 28, 2018 from \tempurl}


\bibitem[\protect\citeauthoryear{hinduexistence.org}{hinduexistence.org}{2012}]%
        {HinduExistence:2012ramu}
\bibfield{author}{\bibinfo{person}{hinduexistence.org}.}
  \bibinfo{year}{2012}\natexlab{}.
\newblock \bibinfo{title}{Buddhist-Hindu temples, homes burned, looted in Ramu,
  Patia… Cox’s Bazar -Bangladesh.}
\newblock
\newblock
\urldef\tempurl%
\url{https://bit.ly/2wsrTb5}
\showURL{%
\tempurl}


\bibitem[\protect\citeauthoryear{hinduexistence.org}{hinduexistence.org}{2016a}]%
        {HinduExistence:2016bbaria}
\bibfield{author}{\bibinfo{person}{hinduexistence.org}.}
  \bibinfo{year}{2016}\natexlab{a}.
\newblock \bibinfo{title}{Brahmanbaria proves Facebook as a Jihadi tool in
  Bangladesh to destroy Hindus!}
\newblock
\newblock
\urldef\tempurl%
\url{https://bit.ly/2MHa5TW}
\showURL{%
\tempurl}


\bibitem[\protect\citeauthoryear{hinduexistence.org}{hinduexistence.org}{2016b}]%
        {HinduExistence:2013pabna}
\bibfield{author}{\bibinfo{person}{hinduexistence.org}.}
  \bibinfo{year}{2016}\natexlab{b}.
\newblock \bibinfo{title}{Dying Hindus in Bangladesh: A concern of US, not of
  India. What a shame!}
\newblock
\newblock
\urldef\tempurl%
\url{http://tiny.cc/k0vmxy}
\showURL{%
\tempurl}


\bibitem[\protect\citeauthoryear{hinduexistence.org}{hinduexistence.org}{2016c}]%
        {HinduExistence:2014comilla}
\bibfield{author}{\bibinfo{person}{hinduexistence.org}.}
  \bibinfo{year}{2016}\natexlab{c}.
\newblock \bibinfo{title}{Hindus again attacked in Bangladesh on false rumours
  of defaming Muhammad in facebook.}
\newblock
\newblock
\urldef\tempurl%
\url{https://bit.ly/2omhppa}
\showURL{%
\tempurl}


\bibitem[\protect\citeauthoryear{hinduexistence.org}{hinduexistence.org}{2017}]%
        {HinduExistence:2017rangpur}
\bibfield{author}{\bibinfo{person}{hinduexistence.org}.}
  \bibinfo{year}{2017}\natexlab{}.
\newblock \bibinfo{title}{Jihadi mob sets fire upon Hindu village in Rangpur
  over rumoured Facebook post from fake account.}
\newblock
\newblock
\urldef\tempurl%
\url{https://bit.ly/2MXmP8d}
\showURL{%
\tempurl}


\bibitem[\protect\citeauthoryear{hinduhumanrights.info}{hinduhumanrights.info}{2013}]%
        {HinduHumanRights:2013false}
\bibfield{author}{\bibinfo{person}{hinduhumanrights.info}.}
  \bibinfo{year}{2013}\natexlab{}.
\newblock \bibinfo{title}{Hindus attacked in Bangladesh over false facebook
  post}.
\newblock
\newblock
\urldef\tempurl%
\url{http://www.hinduhumanrights.info/hindus-attacked-in-bangladesh-over-false-facebook-post/}
\showURL{%
\tempurl}


\bibitem[\protect\citeauthoryear{hindusamhati.net}{hindusamhati.net}{2014}]%
        {HinduSamhati:2014smashed}
\bibfield{author}{\bibinfo{person}{hindusamhati.net}.}
  \bibinfo{year}{2014}\natexlab{}.
\newblock \bibinfo{title}{Hindus Smashed Up In Comilla, Bangladesh}.
\newblock
\newblock
\urldef\tempurl%
\url{http://hindusamhati.net/news/details/109}
\showURL{%
\tempurl}


\bibitem[\protect\citeauthoryear{Janis~Wolak and Finkelhor}{Janis~Wolak and
  Finkelhor}{2007}]%
        {harassment:01}
\bibfield{author}{\bibinfo{person}{Kimberly J.~Mitchell Janis~Wolak, J.D.}
  {and} \bibinfo{person}{David Finkelhor}.} \bibinfo{year}{2007}\natexlab{}.
\newblock \showarticletitle{Does Online Harassment Constitute Bullying? An
  Exploration of Online Harassment by Known Peers and Online-Only Contacts}.
\newblock \bibinfo{journal}{\emph{Adolescent Health}}  \bibinfo{volume}{41}
  (\bibinfo{date}{December} \bibinfo{year}{2007}), \bibinfo{pages}{S51--S58}.
\newblock
Issue 6.
\urldef\tempurl%
\url{https://doi.org/10.1016/j.jadohealth.2007.08.019}
\showDOI{\tempurl}


\bibitem[\protect\citeauthoryear{Kim and Lee}{Kim and Lee}{2011}]%
        {Kim2011TheFP}
\bibfield{author}{\bibinfo{person}{Junghyun Kim} {and}
  \bibinfo{person}{Jong-Eun~Roselyn Lee}.} \bibinfo{year}{2011}\natexlab{}.
\newblock \showarticletitle{The Facebook Paths to Happiness: Effects of the
  Number of Facebook Friends and Self-Presentation on Subjective Well-Being}.
\newblock \bibinfo{journal}{\emph{Cyberpsychology, behavior and social
  networking}}  \bibinfo{volume}{14 6} (\bibinfo{year}{2011}),
  \bibinfo{pages}{359--64}.
\newblock


\bibitem[\protect\citeauthoryear{King, Walpole, and Lamon}{King
  et~al\mbox{.}}{2007}]%
        {King2007SurfAT}
\bibfield{author}{\bibinfo{person}{Jonathan~E King}, \bibinfo{person}{Carolyn~E
  Walpole}, {and} \bibinfo{person}{Kristi Lamon}.}
  \bibinfo{year}{2007}\natexlab{}.
\newblock \showarticletitle{Surf and turf wars online--growing implications of
  Internet gang violence.}
\newblock \bibinfo{journal}{\emph{The Journal of adolescent health : official
  publication of the Society for Adolescent Medicine}}  \bibinfo{volume}{41 6
  Suppl 1} (\bibinfo{year}{2007}), \bibinfo{pages}{S66--8}.
\newblock


\bibitem[\protect\citeauthoryear{M.~Farhad, Rahman~Minar, and
  Majumder}{M.~Farhad et~al\mbox{.}}{2017}]%
        {Minar:2017vital}
\bibfield{author}{\bibinfo{person}{S M.~Farhad}, \bibinfo{person}{Matiur
  Rahman~Minar}, {and} \bibinfo{person}{Sudipta Majumder}.}
  \bibinfo{year}{2017}\natexlab{}.
\newblock \showarticletitle{Measurement of Vital Signs with Non-invasive and
  Wireless Sensing Technologies and Health Monitoring}.
\newblock \bibinfo{journal}{\emph{Journal of Advances in Information
  Technology}} (\bibinfo{date}{01} \bibinfo{year}{2017}),
  \bibinfo{pages}{187--193}.
\newblock
\urldef\tempurl%
\url{https://doi.org/10.12720/jait.8.3.187-193}
\showDOI{\tempurl}


\bibitem[\protect\citeauthoryear{Mazzara, Biselli, Greco, Dragoni, Marraffa,
  Qamar, and de~Nicola}{Mazzara et~al\mbox{.}}{2013}]%
        {MBGDMQN2013}
\bibfield{author}{\bibinfo{person}{Manuel Mazzara}, \bibinfo{person}{Luca
  Biselli}, \bibinfo{person}{Pier~Paolo Greco}, \bibinfo{person}{Nicola
  Dragoni}, \bibinfo{person}{Antonio Marraffa}, \bibinfo{person}{Nafees Qamar},
  {and} \bibinfo{person}{Simona de Nicola}.} \bibinfo{year}{2013}\natexlab{}.
\newblock \bibinfo{booktitle}{\emph{Social networks and collective
  intelligence: a return to the agora}}.
\newblock \bibinfo{publisher}{IGI Global}.
\newblock


\bibitem[\protect\citeauthoryear{Patton, Eschmann, and Butler}{Patton
  et~al\mbox{.}}{2013}]%
        {Patton2013InternetBN}
\bibfield{author}{\bibinfo{person}{Desmond~Upton Patton},
  \bibinfo{person}{Robert~D. Eschmann}, {and} \bibinfo{person}{Dirk~A.
  Butler}.} \bibinfo{year}{2013}\natexlab{}.
\newblock \showarticletitle{Internet banging: New trends in social media, gang
  violence, masculinity and hip hop}.
\newblock \bibinfo{journal}{\emph{Computers in Human Behavior}}
  \bibinfo{volume}{29} (\bibinfo{year}{2013}), \bibinfo{pages}{A54--A59}.
\newblock


\bibitem[\protect\citeauthoryear{Patton, Eschmann, Elsaesser, and
  Bocanegra}{Patton et~al\mbox{.}}{2016}]%
        {Patton2016SticksSA}
\bibfield{author}{\bibinfo{person}{Desmond~Upton Patton},
  \bibinfo{person}{Robert~D. Eschmann}, \bibinfo{person}{Caitlin Elsaesser},
  {and} \bibinfo{person}{Eddie Bocanegra}.} \bibinfo{year}{2016}\natexlab{}.
\newblock \showarticletitle{Sticks, stones and Facebook accounts: What violence
  outreach workers know about social media and urban-based gang violence in
  Chicago}.
\newblock \bibinfo{journal}{\emph{Computers in Human Behavior}}
  \bibinfo{volume}{65} (\bibinfo{year}{2016}), \bibinfo{pages}{591--600}.
\newblock


\bibitem[\protect\citeauthoryear{Paul, Baker, and Cochran}{Paul
  et~al\mbox{.}}{2012}]%
        {Paul:2012:EOS:2365365.2365675}
\bibfield{author}{\bibinfo{person}{Jomon~Aliyas Paul}, \bibinfo{person}{Hope~M.
  Baker}, {and} \bibinfo{person}{Justin~Daniel Cochran}.}
  \bibinfo{year}{2012}\natexlab{}.
\newblock \showarticletitle{Effect of Online Social Networking on Student
  Academic Performance}.
\newblock \bibinfo{journal}{\emph{Comput. Hum. Behav.}} \bibinfo{volume}{28},
  \bibinfo{number}{6} (\bibinfo{date}{Nov.} \bibinfo{year}{2012}),
  \bibinfo{pages}{2117--2127}.
\newblock
\showISSN{0747-5632}
\urldef\tempurl%
\url{https://doi.org/10.1016/j.chb.2012.06.016}
\showDOI{\tempurl}


\bibitem[\protect\citeauthoryear{persecution.in}{persecution.in}{2013}]%
        {Persecution:2013mob}
\bibfield{author}{\bibinfo{person}{persecution.in}.}
  \bibinfo{year}{2013}\natexlab{}.
\newblock \bibinfo{title}{Bangladesh: Muslim mob goes on rampage, vandalizing
  26 Hindu homes, forcing 150 Hindu families to flee on blasphemy rumors}.
\newblock
\newblock
\urldef\tempurl%
\url{http://www.persecution.in/category/location/bonogram-village-santhia-upazila-pabna-district}
\showURL{%
\tempurl}


\bibitem[\protect\citeauthoryear{Risch and Krestel}{Risch and Krestel}{2018}]%
        {W18-4418}
\bibfield{author}{\bibinfo{person}{Julian Risch} {and} \bibinfo{person}{Ralf
  Krestel}.} \bibinfo{year}{2018}\natexlab{}.
\newblock \showarticletitle{Aggression Identification Using Deep Learning and
  Data Augmentation}. In \bibinfo{booktitle}{\emph{Proceedings of the First
  Workshop on Trolling, Aggression and Cyberbullying (TRAC-2018)}}.
  \bibinfo{publisher}{Association for Computational Linguistics},
  \bibinfo{pages}{150--158}.
\newblock
\urldef\tempurl%
\url{http://aclweb.org/anthology/W18-4418}
\showURL{%
\tempurl}


\bibitem[\protect\citeauthoryear{sciencedaily.com}{sciencedaily.com}{2013}]%
        {Gangviolence}
\bibfield{author}{\bibinfo{person}{sciencedaily.com}.}
  \bibinfo{year}{2013}\natexlab{}.
\newblock \bibinfo{title}{Study explores gang activity on the internet}.
\newblock
\newblock
\urldef\tempurl%
\url{https://www.sciencedaily.com/releases/2013/03/130326101525.htm}
\showURL{%
Retrieved August 28, 2018 from \tempurl}


\bibitem[\protect\citeauthoryear{Statista}{Statista}{2017}]%
        {Statista2017}
\bibfield{author}{\bibinfo{person}{Statista}.} \bibinfo{year}{2017}\natexlab{}.
\newblock \bibinfo{title}{Share of women worldwide who have ever experienced
  abuse or harassment on selected websites and social media platforms as of
  July 2017}.
\newblock
\newblock
\urldef\tempurl%
\url{tiny.cc/s43nxy}
\showURL{%
Retrieved August 28, 2018 from \tempurl}


\bibitem[\protect\citeauthoryear{Statista}{Statista}{2018}]%
        {Statista2018}
\bibfield{author}{\bibinfo{person}{Statista}.} \bibinfo{year}{2018}\natexlab{}.
\newblock \bibinfo{title}{Number of monthly active Facebook users worldwide as
  of 2nd quarter 2018 (in millions)}.
\newblock
\newblock
\urldef\tempurl%
\url{tiny.cc/w23nxy}
\showURL{%
Retrieved August 27, 2018 from \tempurl}


\bibitem[\protect\citeauthoryear{Susan~Hopkins}{Susan~Hopkins}{2015}]%
        {domesticviolence}
\bibfield{author}{\bibinfo{person}{Jenny~Ostini Susan~Hopkins}.}
  \bibinfo{year}{2015}\natexlab{}.
\newblock \bibinfo{title}{Domestic violence and Facebook: harassment takes new
  forms in the social media age}.
\newblock
\newblock
\urldef\tempurl%
\url{tiny.cc/r63nxy}
\showURL{%
Retrieved August 28, 2018 from \tempurl}


\bibitem[\protect\citeauthoryear{thedailystar.net}{thedailystar.net}{2012a}]%
        {DailyStar:2012devil}
\bibfield{author}{\bibinfo{person}{thedailystar.net}.}
  \bibinfo{year}{2012}\natexlab{a}.
\newblock \bibinfo{title}{A devil's design}.
\newblock
\newblock
\urldef\tempurl%
\url{http://www.thedailystar.net/news-detail-253751}
\showURL{%
\tempurl}


\bibitem[\protect\citeauthoryear{thedailystar.net}{thedailystar.net}{2012b}]%
        {DailyStar:2012linked}
\bibfield{author}{\bibinfo{person}{thedailystar.net}.}
  \bibinfo{year}{2012}\natexlab{b}.
\newblock \bibinfo{title}{Extremists 'linked'}.
\newblock
\newblock
\urldef\tempurl%
\url{http://www.thedailystar.net/news-detail-251955}
\showURL{%
\tempurl}


\bibitem[\protect\citeauthoryear{thedailystar.net}{thedailystar.net}{2012c}]%
        {DailyStar:2012soul}
\bibfield{author}{\bibinfo{person}{thedailystar.net}.}
  \bibinfo{year}{2012}\natexlab{c}.
\newblock \bibinfo{title}{Tearing out the soul}.
\newblock
\newblock
\urldef\tempurl%
\url{http://www.thedailystar.net/news-detail-252079}
\showURL{%
\tempurl}


\bibitem[\protect\citeauthoryear{thedailystar.net}{thedailystar.net}{2013}]%
        {DailyStar:2013attacked}
\bibfield{author}{\bibinfo{person}{thedailystar.net}.}
  \bibinfo{year}{2013}\natexlab{}.
\newblock \bibinfo{title}{Hindus attacked in Pabna}.
\newblock
\newblock
\urldef\tempurl%
\url{http://www.thedailystar.net/news/hindus-attacked-in-pabna}
\showURL{%
\tempurl}


\bibitem[\protect\citeauthoryear{thedailystar.net}{thedailystar.net}{2014}]%
        {DailyStar:2014rumour}
\bibfield{author}{\bibinfo{person}{thedailystar.net}.}
  \bibinfo{year}{2014}\natexlab{}.
\newblock \bibinfo{title}{Hindu houses attacked on Facebook rumours}.
\newblock
\newblock
\urldef\tempurl%
\url{http://www.thedailystar.net/hindu-houses-attacked-on-facebook-rumours-22087}
\showURL{%
\tempurl}


\bibitem[\protect\citeauthoryear{thedailystar.net}{thedailystar.net}{2016a}]%
        {DailyStar:2016suspect}
\bibfield{author}{\bibinfo{person}{thedailystar.net}.}
  \bibinfo{year}{2016}\natexlab{a}.
\newblock \bibinfo{title}{Nasirnagar attack: Prime suspect on 4-day remand}.
\newblock
\newblock
\urldef\tempurl%
\url{http://www.thedailystar.net/country/prime-suspect-nasirnagar-attack-arrested-1322311}
\showURL{%
\tempurl}


\bibitem[\protect\citeauthoryear{thedailystar.net}{thedailystar.net}{2016b}]%
        {DailyStar:2016planned}
\bibfield{author}{\bibinfo{person}{thedailystar.net}.}
  \bibinfo{year}{2016}\natexlab{b}.
\newblock \bibinfo{title}{Planned attack 'to grab land'}.
\newblock
\newblock
\urldef\tempurl%
\url{http://www.thedailystar.net/frontpage/planned-attack-grab-land-1308718}
\showURL{%
\tempurl}


\bibitem[\protect\citeauthoryear{thedailystar.net}{thedailystar.net}{2016c}]%
        {DailyStar:2016police}
\bibfield{author}{\bibinfo{person}{thedailystar.net}.}
  \bibinfo{year}{2016}\natexlab{c}.
\newblock \bibinfo{title}{Police file two cases in this connection}.
\newblock
\newblock
\urldef\tempurl%
\url{http://www.thedailystar.net/country/attack-hindus-nasrinagar-uno-transferred-1310440}
\showURL{%
\tempurl}


\bibitem[\protect\citeauthoryear{thedailystar.net}{thedailystar.net}{2017a}]%
        {DailyStar:2017clash}
\bibfield{author}{\bibinfo{person}{thedailystar.net}.}
  \bibinfo{year}{2017}\natexlab{a}.
\newblock \bibinfo{title}{Clash over Facebook post, 1 killed in Rangpur}.
\newblock
\newblock
\urldef\tempurl%
\url{https://bit.ly/2PNFpOx}
\showURL{%
\tempurl}


\bibitem[\protect\citeauthoryear{thedailystar.net}{thedailystar.net}{2017b}]%
        {DailyStar:2017mayhem}
\bibfield{author}{\bibinfo{person}{thedailystar.net}.}
  \bibinfo{year}{2017}\natexlab{b}.
\newblock \bibinfo{title}{Mayhem over Facebook post}.
\newblock
\newblock
\urldef\tempurl%
\url{http://www.thedailystar.net/frontpage/mayhem-over-facebook-post-1489402}
\showURL{%
\tempurl}


\bibitem[\protect\citeauthoryear{thedailystar.net}{thedailystar.net}{2017c}]%
        {DailyStar:2017accused}
\bibfield{author}{\bibinfo{person}{thedailystar.net}.}
  \bibinfo{year}{2017}\natexlab{c}.
\newblock \bibinfo{title}{Thakurpara Mayhem: Accused Tito Chandra Roy held in
  Nilphamari}.
\newblock
\newblock
\urldef\tempurl%
\url{https://bit.ly/2PPI3Dh}
\showURL{%
\tempurl}


\bibitem[\protect\citeauthoryear{thedailystar.net}{thedailystar.net}{2017d}]%
        {DailyStar:2017framed}
\bibfield{author}{\bibinfo{person}{thedailystar.net}.}
  \bibinfo{year}{2017}\natexlab{d}.
\newblock \bibinfo{title}{Thakurpara Mayhem: Influential locals spread hatred}.
\newblock
\newblock
\urldef\tempurl%
\url{http://www.thedailystar.net/backpage/thakurpara-hindu-was-framed-1490338}
\showURL{%
\tempurl}


\bibitem[\protect\citeauthoryear{Tripathi}{Tripathi}{2017}]%
        {harassment:04}
\bibfield{author}{\bibinfo{person}{Vivek Tripathi}.}
  \bibinfo{year}{2017}\natexlab{}.
\newblock \showarticletitle{Youth Violence and Social Media}.
\newblock \bibinfo{journal}{\emph{Social Sciences}}  \bibinfo{volume}{52}
  (\bibinfo{date}{August} \bibinfo{year}{2017}).
\newblock
Issue 1-3.
\urldef\tempurl%
\url{https://doi.org/10.1080/09718923.2017.1352614}
\showDOI{\tempurl}


\bibitem[\protect\citeauthoryear{Wikipedia}{Wikipedia}{2017a}]%
        {Wikipedia:2012ramu}
\bibfield{author}{\bibinfo{person}{Wikipedia}.}
  \bibinfo{year}{2017}\natexlab{a}.
\newblock \bibinfo{title}{2012 Ramu violence}.
\newblock
\newblock
\urldef\tempurl%
\url{https://en.wikipedia.org/wiki/2012_Ramu_violence}
\showURL{%
\tempurl}


\bibitem[\protect\citeauthoryear{Wikipedia}{Wikipedia}{2017b}]%
        {Facebook:2017wk}
\bibfield{author}{\bibinfo{person}{Wikipedia}.}
  \bibinfo{year}{2017}\natexlab{b}.
\newblock \bibinfo{title}{Facebook}.
\newblock
\newblock
\urldef\tempurl%
\url{https://en.wikipedia.org/wiki/Facebook}
\showURL{%
Retrieved August 27, 2018 from \tempurl}


\bibitem[\protect\citeauthoryear{Wikipedia}{Wikipedia}{2017c}]%
        {Wikipedia:2017ramu}
\bibfield{author}{\bibinfo{person}{Wikipedia}.}
  \bibinfo{year}{2017}\natexlab{c}.
\newblock \bibinfo{title}{Ramu, Cox's Bazar}.
\newblock
\newblock
\urldef\tempurl%
\url{https://en.wikipedia.org/wiki/Ramu,_Cox%27s_Bazar}
\showURL{%
\tempurl}


\bibitem[\protect\citeauthoryear{Wikipedia}{Wikipedia}{2017d}]%
        {Wikipedia:2017upazilla}
\bibfield{author}{\bibinfo{person}{Wikipedia}.}
  \bibinfo{year}{2017}\natexlab{d}.
\newblock \bibinfo{title}{Ramu Upazila}.
\newblock
\newblock
\urldef\tempurl%
\url{https://en.wikipedia.org/wiki/Ramu_Upazila}
\showURL{%
\tempurl}


\bibitem[\protect\citeauthoryear{wired.com}{wired.com}{2017}]%
        {Wired:2017suicide}
\bibfield{author}{\bibinfo{person}{wired.com}.}
  \bibinfo{year}{2017}\natexlab{}.
\newblock \bibinfo{title}{ARTIFICIAL INTELLIGENCE IS LEARNING TO PREDICT AND
  PREVENT SUICIDE}.
\newblock
\newblock
\urldef\tempurl%
\url{https://www.wired.com/2017/03/artificial-intelligence-learning-predict-prevent-suicide/}
\showURL{%
\tempurl}


\end{thebibliography}

\end{document}